# Atomic-scale tracking of topological defect motion and incommensurate charge order melting

Noah Schnitzer[1], Berit H. Goodge[2,3,*], Gregory Powers[2], Jaewook Kim[4,5], Sang-Wook Cheong[4,5], Ismail El Baggari[6,7], and Lena F. Kourkoutis[2,8]

[1] Department of Materials Science and Engineering, Cornell University, Ithaca NY 14853 USA
[2] School of Applied and Engineering Physics, Cornell University, Ithaca NY 14853, USA
[3] Max Planck Institute for Chemical Physics of Solids, Dresden 01187, Germany
[4] Department of Physics and Astronomy, Rutgers University, Piscataway NJ 08854, USA
[5] Rutgers Center for Emergent Materials, Rutgers University, Piscataway NJ 08854, USA
[6] Department of Physics, Cornell University, Ithaca NY 14853 USA
[7] Rowland Institute at Harvard, Harvard University, Cambridge MA 02142, USA
[8] Kavli Institute at Cornell for Nanoscale Science, Cornell University, Ithaca NY 14853, USA
[*] e-mail: Berit.Goodge@cpfs.mpg.de

## Abstract
Charge order pervades the phase diagrams of quantum materials where it competes with superconducting and magnetic phases, hosts electronic phase transitions and topological defects, and couples to the lattice generating intricate structural distortions. Incommensurate charge order is readily stabilized in manganese oxides where it is associated with anomalous electronic and magnetic properties, but its nanoscale structural inhomogeneity complicates precise characterization and understanding of its relationship with competing phases. Leveraging atomic-resolution variable temperature cryogenic scanning transmission electron microscopy, we characterize the thermal evolution of charge order as it transforms from its ground state in a model manganite system. We find that mobile networks of discommensurations and dislocations generate phase inhomogeneity and induce global incommensurability in an otherwise lattice-locked modulation. Driving the order to melt at high temperatures, the discommensuration density grows and regions of order locally decouple from the lattice periodicity.

## Introduction
Strong electron interactions can give rise to exotic behavior such as superconductivity and charge density waves, but also host disorder and are disrupted with increasing temperature or under applied fields. Properties of both fundamental and technological interest in strongly correlated materials arise not just in clean single-phase systems, but also near phase boundaries where competing orders interact in states characterized by nanoscale inhomogeneity [1–5]. Thus, achieving understanding and control of strongly correlated phenomena requires probing these materials beyond their ground states as phases coalesce, evolve, and melt at the boundary between order and disorder.

In these conditions, spatial inhomogeneity manifests through a variety of active micro- and nanoscale features, such as phase coexistence [6–9], domain structures [10,11], and topological defects [12–17]. Topological defects are a particularly interesting case: they underly disparate phenomena such as liquid crystallinity, plasticity, turbulent flow in active matter, and early universe dynamics [18–22]. The presence of these defects not only results in the local suppression of order, but can also alter long-range correlations with profound consequences on macroscopic properties and the nature of phase transitions [14,23–29]. Observing how these features modulate electronic order and evolve under external stimulus – for instance, how topological defects move, bind, unbind, and disrupt order – is of fundamental importance, but nanoscale structural inhomogeneity poses a significant challenge to both theoretical understanding and ensemble characterization approaches.

Charge order, a symmetry-breaking modulation of electron density, is a prototypical example of strongly correlated electronic behavior which emerges in a broad set of materials with varying degrees of coupling to the atomic lattice [4,30,31]. At the weak coupling limit, Fermi surface nesting can give rise to charge density waves, as found, for example in $2H$-$TaSe_2$. In the strongly coupled extreme, on the other hand, real-space patterns of positive charge (holes) generate charge ordered stripes in a variety of manganese, copper, and nickel oxides [30]. These states are remarkable for their rich interactions with other electronic and magnetic orders: for instance, the intertwining of charge density waves and superconductivity gives rise to pair density waves, and colossal magnetoresistance emerges from nanoscale phase coexistence driven by competition between charge and magnetic order [2,6,32–36].

Great effort has focused on the emergence of incommensurate charge order, with a wavevector not locked to integer units of the atomic lattice, and its association with topological defects. The pioneering theory of W.L. McMillan predicted that in charge density waves weakly coupled to the lattice, ensemble average measurements of incommensurate wavevectors arise from the presence of localized phase slips, or discommensurations, in the order with a density proportional to the incommensurability [37]. This was confirmed by dark-field transmission electron microscopy (TEM) investigations which also revealed dislocations in the charge density waves at points where the dislocation lines meet, likewise in agreement with McMillan's theory and firmly establishing the role of topological defects in forming incommensurate charge density waves [38,39]. Intriguingly, discommensurations have also been imaged in the charge ordered stripes of doped holes in the manganites – perovskite oxides of the form $AMnO_3$ – suggesting the same mechanism might underly incommensurability in both the weak- and strong-coupling regimes [40,41,6,42]. Additional studies identified the presence of dislocations in charge order stripes, found the incommensurability of the charge order is tuned by the hole doping, and measured changes in incommensurability associated with melting into the high temperature charge disordered phase [29,43,44]. This behavior makes manganites an ideal model to study both the origins of incommensurate charge order in the strongly-coupled regime and its evolution amidst order parameter fluctuations as it departs the ordered state.

Simultaneously achieving the high spatial resolution necessary to resolve how nanoscale features govern structure and properties with variable temperature control to drive electronic transitions in charge ordered systems with presents a major *in situ* characterization challenge. While many scattering and optical probes (e.g. X-ray diffraction, near-field scanning optical microscopy) as well as bulk transport measurements allow continuous temperature control, they lack the spatial resolution to probe the relevant atomic-scale behavior. On the other hand, scanning probe techniques such as scanning tunneling microscopy can provide sufficient spatial resolution but are sensitive only to certain surfaces. Scanning transmission electron microscopy (STEM) measures the complete projected structure of materials with atomic resolution, enabling the picometer scale atomic displacements arising from charge order to be visualized in bulk materials, at interfaces, and near defects [43]. In recent years, the difficulties in achieving the high mechanical stability required for STEM imaging in the presence of a cryogen have been met, establishing a nascent capability to visualize strongly correlated phenomena at the low temperatures where they often emerge [45–53].

Thus far, however, atomic resolution cryogenic STEM (cryo-STEM) has been limited to fixed temperatures around ~100 K, set by the thermal equilibrium with the liquid nitrogen cryogen. This constraint arises from the design of conventional monolithic cryo-STEM specimen holders which cannot heat the sample without heating the entire device, thereby upsetting the thermal equilibrium and generating significant mechanical instabilities including large sample drift which make acquiring high resolution images suitable for quantitative analysis impossible. Now, a new generation of cryogenic specimen holders integrating micro-electromechanical systems (MEMS) can locally heat the sample without substantively affecting the thermal equilibrium of the system or upsetting the cryogen [54]. Here, we combine liquid nitrogen cooling to achieve a low-temperature baseline with MEMS-based restive heating to achieve control of the sample temperature over the range of ~115-1000 K with the exacting stability required for quantitative atomic resolution STEM imaging. We leverage this breakthrough experimental capability to report the first atomic-scale measurements at identical locations across variable cryogenic temperatures to track local charge order fluctuations and topological defects, revealing two distinct regimes of temperature dependent behavior in a mixed-valence manganite. We directly visualize the quasi-static behavior of the topological defects at low temperatures, which accommodate global incommensurability in an otherwise locally commensurate lattice-locked order. At high temperature, we track the increasing density of discommensurations and dislocations as the order melts, revealing how nanoscale structural transformation drives changes in ensemble parameters.

## Results
### Periodic Lattice Displacements

Charge ordered $Bi_{1-x}Sr_{x-y}Ca_yMnO_3$ single crystals were grown with the flux method with an approximate composition of $x \approx 0.65$ and $y \approx 0.47$ and crystallized into the *Pnma* distorted perovskite structure, as described previously [43,45]. This composition presents an ideal model system for studying charge order as it lacks the magnetic transitions that complicate and mask the behavior of pure charge order phenomena. At $x$=0.65, the charge order is predicted to be incommensurate with wavevector $\mathbf{q} = (1 - x)\,\mathbf{a}^* = 0.35\,\mathbf{a}^*$ by a previously established empirical relationship, $\mathbf{a}^*$ is the reciprocal lattice vector [44]. Figure 1a shows an atomic-resolution high-angle annular dark-field (HAADF) STEM image where the cation atomic columns are readily visible. Leveraging recent advances in cryo-STEM instrumentation [54], we are able to image with atomic resolution across variable cryogenic temperatures: the image analyzed in Figure 1 is acquired at ~198 K, significantly above the conventional cryo-STEM baseline of ~100 K but far below the onset of the (near room temperature) melting transition. Further information about temperature control is provided in Supplementary Note 1 [55].

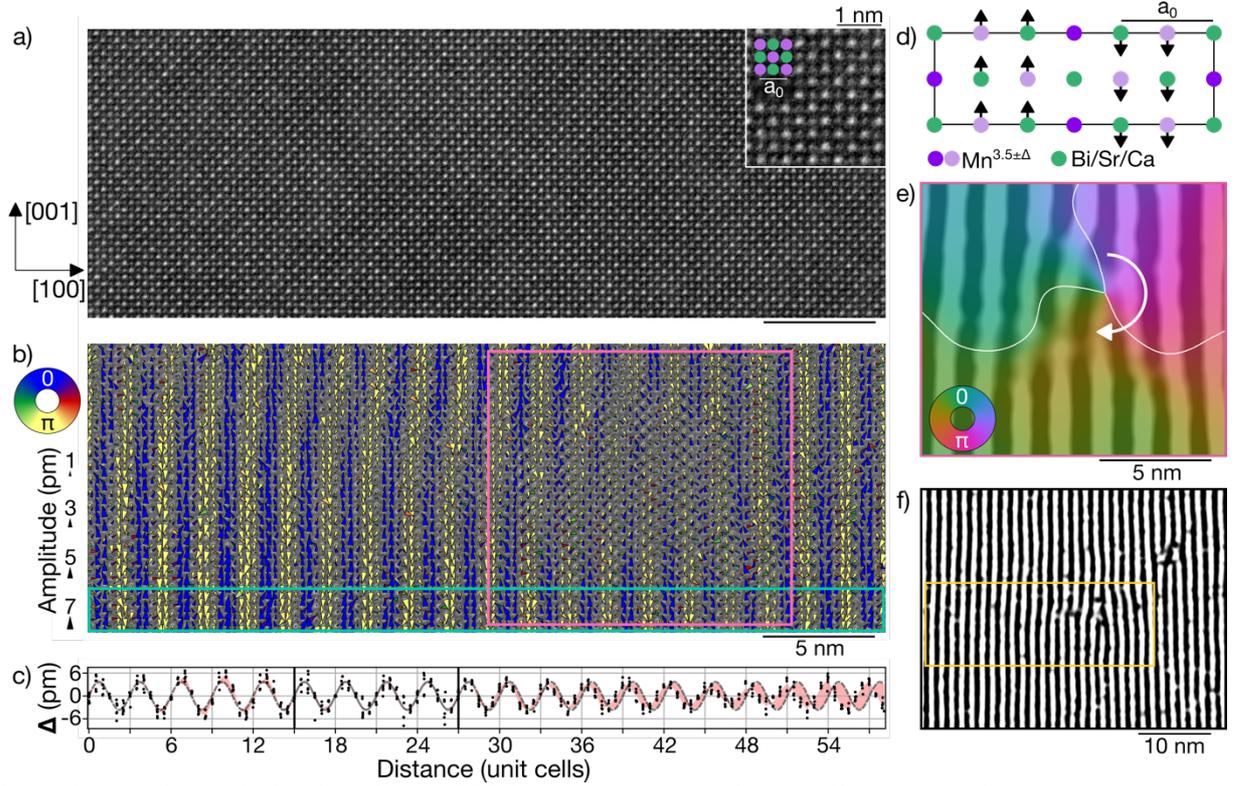

**Figure 1. Atomic resolution imaging and PLD measurement at intermediate cryogenic temperature. a)** Atomic-resolution HAADF-STEM image acquired at 198 K, cropped from a larger field of view. Inset shows an expanded small area where bright shared (Bi,Sr,Ca) sites (green) and less intense Mn sites (purple) can be seen. **b)** Map of PLD displacements in same field of view as (a), represented by arrows with sizes corresponding to the displacement amplitude and colors to the displacement direction. **c)** Black dots mark displacement amplitudes profiled along the [100] direction from the area indicated by a cyan box in (b). A sine wave fit to a commensurate area (gray solid line) is extrapolated over the rest of the profile (gray dashed line) to visualize the change in periodicity around the topological defect. The deviation from this fit (i.e. discommensuration) is shaded in red. d) Displacement of cation sites for a commensurate 3x1 unit cell per the Wigner crystal model. Inequivalent Mn sites are shown in dark and light purple, mixed A-site cation columns are shown in green, with displacements relative to the high symmetry positions shown in black (not to scale). e) A map of the orientation of the PLD stripes from the area indicated by pink rectangle in (b) around a topological defect. Stripes of displacements along [001] are shown in white, displacements along [00$\bar{1}$] are black. Stripes are overlaid with the phase $\phi(\mathbf{r})$ colored as indicated by the wheel in the bottom left, $2\pi/3$ phase contours are overlaid in white. f) A larger field of view map of the PLD stripes, the area shown in (a,b) is boxed in yellow.

Coupling between the charge order and lattice generates picometer-scale periodic lattice displacements (PLDs) of the cations in the *a-c* plane. The PLDs are too small to see by eye in the image and must be extracted through quantitative image analysis [43]. Figure 1b visualizes these displacements relative to an undistorted reference lattice, revealing alternating transverse distortions that form a striped arrangement in real space. A profile of the displacements along the [100] direction (Fig. 1c) shows they are roughly sinusoidal with an approximately 3$a_0$ unit-cell wavelength, where $a_0$ denotes the *Pnma* parent lattice parameter. Accordingly, we parameterize the displacements, $\mathbf{\Delta}(\mathbf{r})$ as:

$$\mathbf{\Delta}(\mathbf{r}) = \mathbf{A}(\mathbf{r}) \sin(\mathbf{q} \cdot \mathbf{r} + \phi(\mathbf{r}))$$

where $\mathbf{A}(\mathbf{r})$ is the displacement amplitude vector, $\phi(\mathbf{r})$ the phase, and $\mathbf{r}$ the position in real space. As we will demonstrate, these components vary in space with profound consequence on the macroscopic behavior of the charge order. Accordingly, we interpret $\mathbf{\Delta}(\mathbf{r})$ as a structural order parameter of the charge ordering.

Rather than developing a uniformly incommensurate periodicity, we find that large areas of the material host lattice-locked, commensurate order with $\mathbf{q} = \frac{1}{3}\mathbf{a^*}$ as shown schematically in Figure 1d and in agreement with the "Wigner crystal" model previously refined in $La_{1/3}Ca_{2/3}MnO_3$ (see also Supp. Fig. S3) [43,45,55]. These commensurate regions, however, are sparsely interspersed with defects in the charge order superlattice which, like disclinations and dislocations in liquid crystals, break symmetry altering the nature of the order. Figure 1b shows one such edge dislocation-like topological defect visible in the cation displacements and emphasized in a map showing the stripes formed by thresholding the alternating cation displacement directions (Fig. 1e). McMillan's theory of discommensurations in charge density waves suggests that such dislocations in the charge order emerge from the termination of three discommensurations, each imparting a $2\pi/3$ phase slip, resulting in a full $2\pi$ winding and the insertion of an additional half plane [37]. Overlaid on Figure 1e is the spatially varying phase, $\phi(\mathbf{r})$, of the displacements, which offers rich information about the topology of the order [56], visualizing the $2\pi$ winding of the phase at the junction of three discommensurations at the defect core. Interestingly, although the stripe order in this material represents the strong-coupling limit of charge modulations, the same dislocations and discommensurations found in weakly coupled charge density waves permeate the system and are intrinsic to the charge order – no corresponding crystalline defect is present in the atomic lattice (Fig. 1a, Supp. Fig. S14). They locally modulate the periodicity of the order, as shown by the deviation of the displacements from a sinusoid fit to a commensurate part of the displacement profile (Fig. 1c). These variations in the periodicity and phase of the order induced by the defects drive the incommensurability of the order, as previously characterized in cuprates and manganites, and demonstrate the applicability of McMillan's discommensuration theory to the strongly coupled regime [37,45,57]. A larger field of view showing the mostly ordered, commensurate PLD stripes surrounding the defect is shown in Figure 1f. Details on how these stripe maps, the spatially varying phase, and other features of the order are extracted from HAADF-STEM images are provided in Supplementary Note 2 [55].

**Thermal Evolution of the Order**
We explore the evolution of the charge order and its coupling to the lattice as a function of temperature by heating through the order's melting, near room temperature. Charge order correlations averaged over micron scales can first be approached with scattering techniques such as selected area electron diffraction (SAED), which measure the ensemble reciprocal space structure of the material over many unit cells [42]. SAED on the [010] zone axis shows superlattice reflections arising from two orthogonal twins of the transverse unidirectional PLDs. Previous measurements have shown that the twins are anti-correlated in the (010) plane [43]; here we confine our analysis to the modulation along [100] and further high-resolution characterization is performed in regions away from twin domain boundaries where only one unidirectional modulation is present. Figure 2a shows a SAED pattern with cutouts of the (200) Bragg peak and surrounding superlattice reflections as the sample is heated *in situ*. At high temperatures, the superlattice peaks gradually disappear as the order melts. This can be seen more clearly in line profiles through the superlattice peaks across a series of temperatures (Fig. 2b), showing a gradual broadening and loss of intensity beginning between 248 and 273 K until the peak is completely suppressed by 323 K. The wavevector of the peak also gradually reduces over a similar temperature range (Fig. 2c). The full temperature cycle is shown in Supplementary Figure S4 [55]. Although this ensemble view suggests the structure is fairly static at temperatures below the onset of the melting transition, it offers no information about its real-space distribution or potential dynamics thereof. Likewise, while the coupled changes in peak width (associated with the coherence length of the order) and wavevector (periodicity of the order), suggest that these parameters are linked, without real-space sensitivity this limited picture of the melting transition from scattering measurements is insufficient to identify the microscopic mechanism connecting them. For that, it is necessary to go beyond the ensemble and consider the nanoscale details of the system as a function of temperature.

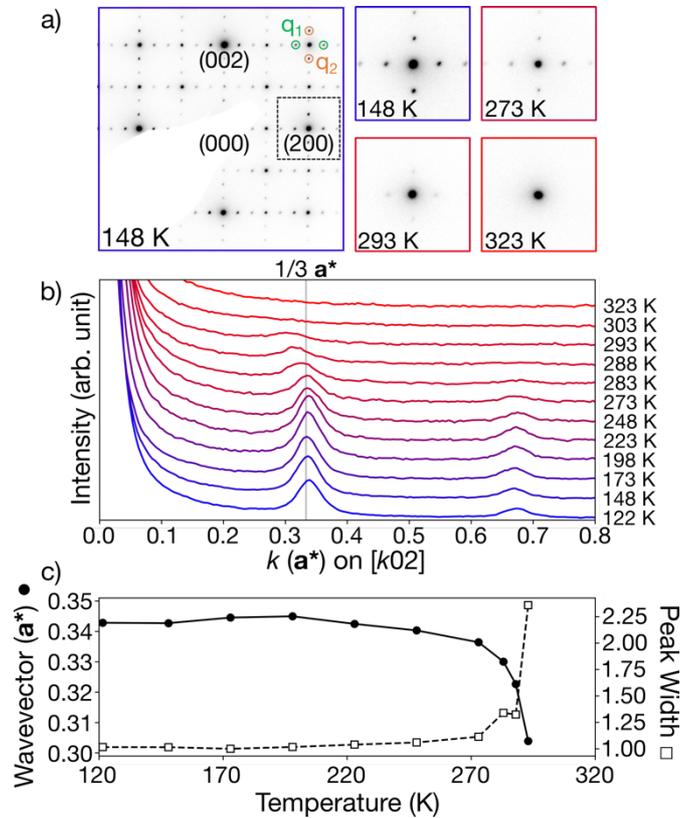

**Figure 2. Ensemble measurement of charge order melting with electron diffraction.** a) A SAED pattern acquired at 148 K on the [010] zone axis, two orthogonal sets of superlattice satellite peaks from PLDs are marked in orange and green. Kinematically forbidden reflections (e.g. (100) and (010)) appear as a result of multiple scattering. Panels on the right show the area marked with a dotted box at 148 K, 273 K, 293 K, and 323 K as the superlattice peaks gradually disappear with heating. Contrast is inverted for visualization. b) Line profiles from diffraction patterns acquired at a series of temperatures along the [100] direction beginning at the (002) Bragg peak. Profiles are normalized to Bragg peak intensity and vertically offset for visualization. Temperature steps between profiles are non-uniform. Commensurate $k = 1/3$ **a*** line added as a guide for the eye. c) Line plots of superlattice peak wavevector (solid line, filled circle markers) and peak width (dashed line, open square markers) as a function of temperature. Peak widths are normalized to the lowest temperature (122 K) width. Peaks above 288 K could not be reliably fit due to low intensity and are not shown.

We begin by considering correlation functions of the spatially varying phase $\phi(\mathbf{r})$ of the PLD, a critical parameter of the charge order extracted from atomic resolution STEM images but lost in SAED. These measurements are acquired spanning an 88x88 nm$^2$ field of view – two orders of magnitude smaller than the SAED sampling area. The autocorrelation of $\phi(\mathbf{r})$ encodes the length scale (**r**) over which the PLD phase is coherent: values close to one indicate the phase is uniform while values near zero mean the local phase is varying or uncorrelated. Radial profiles of PLD phase autocorrelations were calculated from identical regions of the sample through a heating and cooling cycle, showing changes in the correlation length as a function of temperature (Fig. 3a, details in Supplementary Note 3) [55]. The length scale over which the radial profiles decay towards zero measures the correlation length of the charge order, and thus show a very similar trend to the satellite peak width measured with electron diffraction: at low temperatures the correlation functions are almost constant across temperature steps, but near room temperature the length scale at which they begin to drop decreases until our measurement of the order is indistinguishable from noise. The coherent length of the charge order derived from the autocorrelations is plotted alongside that estimated from the SAED satellite peak widths in Supplementary Figure S16, see Supplementary Note 4 for considerations due to the different length scales sampled. Upon cooling back down the order gradually recovers with significant hysteresis until it is almost fully recovered by 131 K. This agreement with the SAED result confirms that the reduced length scale measured here is sufficient to capture the behavior of the order with nanoscale sensitivity, and the melting is associated with reduced phase coherence.

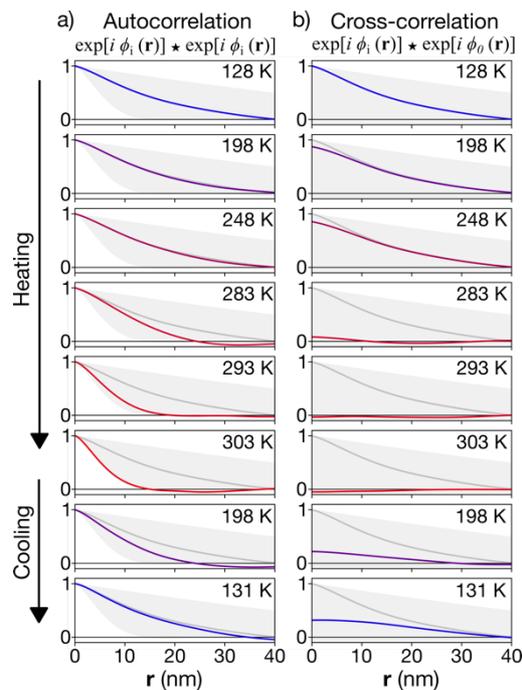

**Figure 3. Correlation functions of PLD phase through a heating-cooling cycle.** a) Radial profiles of autocorrelations of the PLD phase extracted from 88x88 nm$^2$ field of view STEM images. b) Radial profiles of the cross-correlations of the PLD phase extracted from each STEM image with the initial 128 K phase. The gray shaded region represents the range between uncorrelated noise and the maximal correlation for the finite field of view analyzed. Gray solid lines show the initial 128 K autocorrelations for comparison. Details on correlation function calculations are provided in the Supplementary Note 3 [55].

Across the same image series, we also cross-correlate the initial low temperature $\phi(\mathbf{r})$ measured from the same region of interest at each subsequent temperature step (Fig. 3b). This correlation function reflects the spatial coherence of the charge order, or how much the order structurally reconfigures itself – including features like discommensurations and topological defects – as a function of temperature. A value near one indicates the charge order is spatially similar to the initial structure, while a value near zero means the PLD arrangement has been significantly altered. Unlike the autocorrelations, the cross-correlations drop in the first temperature steps. The drop is slight, suggesting that fluctuations in the order are small – potentially due to weak pinning to disorder – but indicates that minor reconfigurations of the stripe order do occur with temperature changes. As the order begins to melt at higher temperatures, the cross-correlation rapidly collapses entirely and does not fully recover upon cooling back down. This indicates that: 1) the structure is in fact not static at temperatures below the melting transition; 2) the melting is associated with a large-scale reconfiguration even before the order is fully suppressed; and 3) the order adopts a distinct arrangement after temperature cycling (see Supplemental Figure S5 for real-space mapping of $\phi(\mathbf{r})$ across the complete heating-cooling cycle, and Supplemental Figure S6 for comparison of charge order stripe arrangements at low temperatures before and after the cycle).

**Real-space mapping of nanoscale structural changes**
Further insight relies upon direct visualization of the real-space charge order arrangement to uncover the mechanisms driving relatively minor reconfigurations at sub-melting temperatures and major reconfigurations accompanying the melting. First, we evaluate the low-temperature minor reconfigurations: which structural features are changing, and why is there no measurable change in the ensemble order? A $\phi(\mathbf{r})$-map of at 116 K over a comparable field of view to the correlation analysis is shown in Figure 4a. Flat phase regions are separated by discommensurations (or phase slips, roughly indicated by the $2\pi/3$ contours) and a network of topological defects with full $2\pi$ phase windings formed at points where three discommensuration lines meet [37]. Figure 4b shows a corresponding map of the stripes described by the PLD orientations. In agreement with previous work, close inspection of the PLD stripe maps reveals that the discommensurations locally alter the PLD wavelength $\lambda = 1/q$, while the phase windings generate dislocations with an additional half-stripe with an orientation set by the winding direction [37,43,45]. These defects locally disrupt the order and impart a change in the average wavevector, accommodating the doping pressure for incommensurability (Supp. Fig. S8) [12,55]. Because the incommensurability is proportional to the discommensuration density and the crystal studied here is nearly commensurate, the defects are well separated by ordered, lattice locked areas in agreement with the relatively large coherence lengths suggested by the phase autocorrelations and diffraction data.

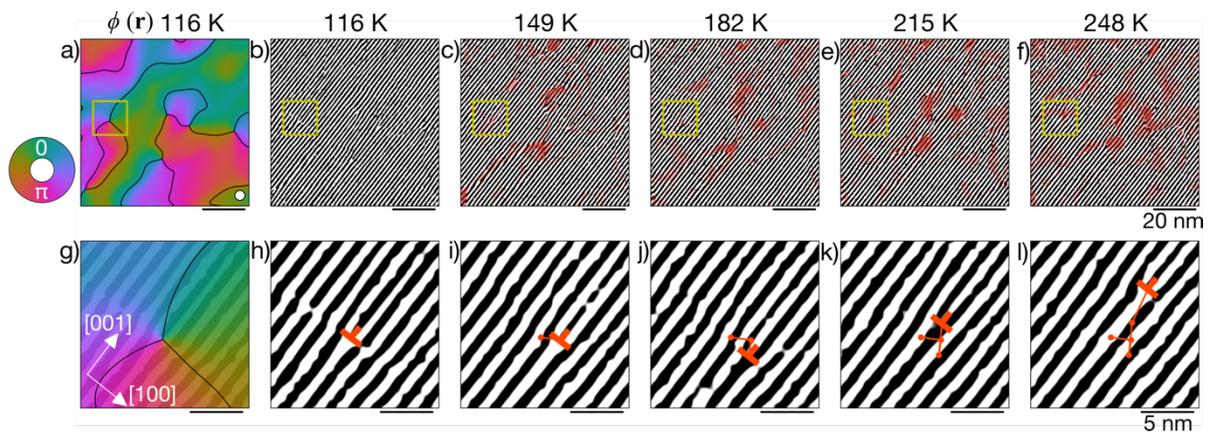

**Figure 4. Charge order defects and motion at low temperatures.** a) Map of the PLD phase across a 77x77 nm$^2$ field of view, overlayed with contours at each $2\pi/3$ front. Coarsening length is denoted by the white circle in the bottom right of the image. b-f) Maps of PLD stripes from precisely the same field of view at (b) 116 K, (c) 149 K, (d) 182 K, (e) 215 K, (f) 248 K. Areas where the PLD stripe positions changed compared to the previous temperature step are highlighted in red. Corresponding phase maps are shown in Supp. Fig. S7. g-l) Expanded views of the region marked with yellow boxes in (a-f). (g) is additionally overlaid with the 116 K stripe map showing a dislocation in the stripes at the position of the phase winding. The position of this topological defect (red T marker) and its trajectory through the material (red line) is marked at each temperature step in (h-l). Scale bars are the same in (a-f) and (g-l).

To explore the temperature-dependent behavior of these defects, maps of the PLD stripes as the material is heated to sub-melting temperatures are shown in Figure 4c-f with red highlighting where the stripe pattern changes compared to the previous temperature step. Remarkably, most of the stripes are almost completely static across the five temperature steps, while changes are localized to the areas immediately surrounding the dislocations and discommensurations, which travel through the lattice with heating. A region of interest around one such topological defect marked with a yellow box is expanded in Figure 4g-l. The overlaid phase and stripe maps show the winding associated with the defect (Fig. 4g), which moves across the field of view on a non-linear trajectory as the sample is heated. Notably, the stripes just nanometers away from the defect core remain practically unchanged (Supp. Fig. S9) [55]. These findings suggest that in this low-temperature, well-ordered regime the charge order is strongly coupled to the lattice except for the areas immediately surrounding the defects in the order. The defects are unable to freely shear through or globally rearrange the surrounding coherent order, and consequently the observed structural changes in this temperature regime are small.

Heating to higher temperatures, we next investigate the structural changes as the order melts. Figure 5a shows a series of stripe maps heating through the melting transition, weighted with the local displacement amplitude to reflect the increasing fluctuations of the PLD amplitude. Corresponding maps of the local amplitude and wavelength are shown in Figure 5b and 5c, histograms are shown in Supplementary Figure S10, and maps of the phase through the temperature cycle are shown in Supplementary Figure S5 [55]. Heating through 273 K the stripes are largely well ordered, showing only motion of isolated defects similar to that in Figure 4. At 283 K, additional discommensurations and dislocations begin to nucleate, creating channels of suppressed amplitude, altering the wavelength of the surrounding PLDs, and disrupting the long-range order. At 293 K, further generation of defects breaks the stripe order into small domains of finite but reduced amplitude, and in almost all remaining ordered regions the stripe wavelength is significantly larger than the original 3 unit-cell commensurate period. This growing defect density drives the concomitant reduction in long range order and increase in incommensurability observed through melting with SAED here and in previous studies of 2/3 hole-doped manganites [42,45]. Heating further to 303 K, the charge order is almost entirely melted with the PLD amplitude below the STEM measurement sensitivity over most of the field of view, and superlattice peaks are no longer visible in the corresponding Fast Fourier Transform (Supp. Fig. S15). Heatmaps showing the increasing extent of reconfigurations with temperature are shown in Supplementary Figure S11 [55].

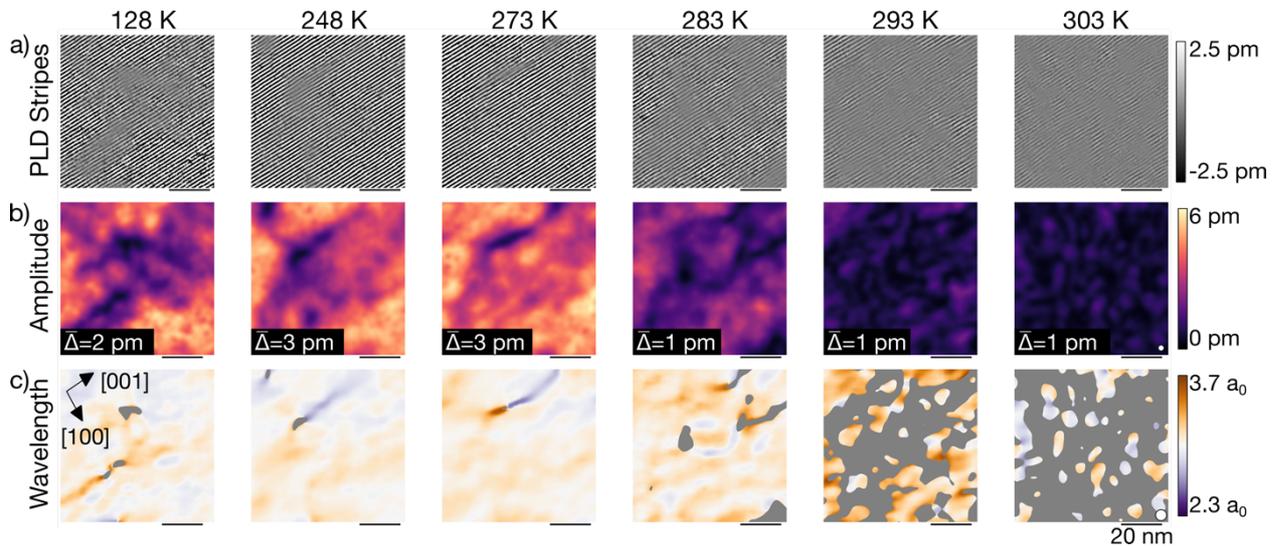

**Figure 5. Real space mapping of charge order melting.** a) Maps of PLD stripes extracted from images acquired at six temperature steps between 128 and 303 K over the same 75x75 nm$^2$ field of view showing a gradual loss of order. b) Maps of the projected PLD amplitude from the same acquisitions as (a). The maps were blurred to average over the PLDs' sinusoidal amplitude variation. Average amplitudes at each temperature step are inset in the lower left of each frame. c) Maps of the PLD wavelength from the same acquisitions as (a). Orange (purple) areas indicate locally longer (shorter) wavelengths and shorter (longer) wavevectors than the commensurate three-unit cell wavelength. Areas where the PLD amplitude fall below the limit of our measurement are excluded from the map and marked in gray.

A key feature observed in the midst of the melting is a decoupling of the charge order from the lattice periodicity. As discussed above, although at low temperatures the average wavevector is slightly larger than the commensurate $q=1/3$ **a*** value, much of the order in fact exhibits a commensurate 3 unit cell wavelength ($\lambda = 3\ a_0$) because the incommensurability is accommodated by local defects. By contrast, at higher temperatures, as the order melts and its amplitude is suppressed to the edge of the precision of measurement with STEM, a growing proportion of the order is truly locally incommensurate with the atomic lattice. Figure 6 shows detailed analysis of a high-amplitude region of partially melted order at 293 K. Similar to observations made at lower temperatures, the PLD stripes are bent and disrupted by defects, but now the defects are separated by 10 nm or less (Fig. 6a). They continue to modulate the periodicity of surrounding stripes, driving them almost uniformly to larger incommensurate values ($\lambda > 3\ a_0$) as shown in a map of the local PLD wavelength (Fig. 6b). Due to the large defect density these incommensurate regions occupy in total a much larger fraction of the field of view than those brought about by the isolated defects found at lower temperatures (e.g., at 128 and 248 K in Fig. 5c), but the defects also prevent the order from forming a coherent, uniformly incommensurate wave. Corresponding cation column displacements and $\phi(\mathbf{r})$ map are shown in Supplementary Figure S12 [55].

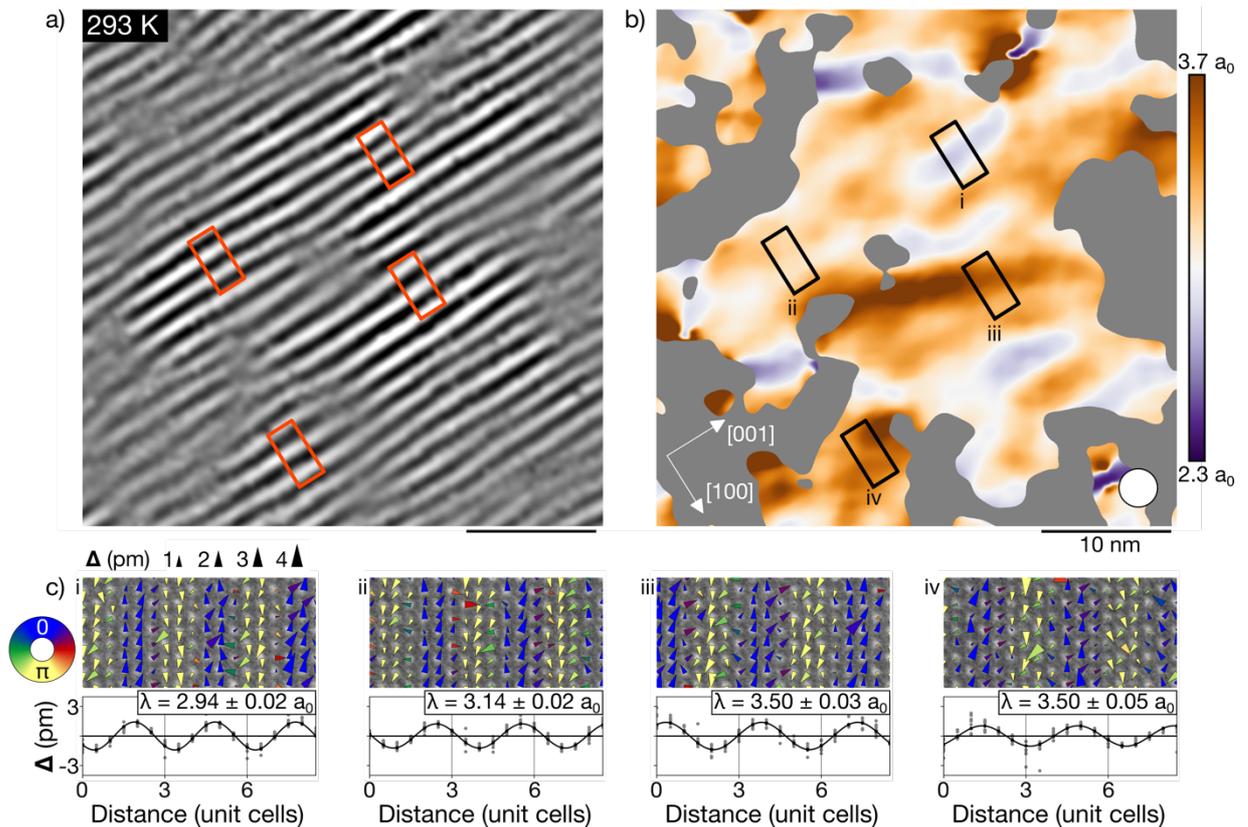

**Figure 6. Nanoscale features of melting charge order.** a) PLD stripes mapped over a 40x40 nm² field of view at 293 K, scaled with local PLD amplitude. b) Coarse grained measurement of PLD wavelength over the same field of view as (a). Areas where the PLD amplitude fall below the limit of our measurement are excluded from the map and marked in gray. c) Expanded PLD maps of four regions (a) and (b), and corresponding displacement profiles (gray dots) overlaid with sine wave fits (black lines).

The continuous variations in wavelength visible in the stripe and wavelength maps indicate that the order also does not lock into a mixture of commensurate periodicities to accommodate the incommensurability. Profiles of the PLDs from selected regions show that the displacements maintain sinusoidal forms but with a range of locally commensurate (or nearly commensurate) and incommensurate wavelengths (Fig. 6c) [29,58]. Together, these results illustrate a regime of temperature-dependent correlations in the charge order distinct from the low temperature incommensurate order. Just as at lower temperatures, charge order incommensurability is driven by defects which modulate the phase of the surrounding order. As the order melts, however, the increasing density of the defect network gives rise to an overall reduction in the charge order wavevector (increase in wavelength) and reduction in coherence length.

**Discussion**

Here, we have investigated the nanoscale thermal evolution of incommensurate charge order, visualizing with atomic resolution its lattice response from ground state to melting. We find topological defects actively shape the microscopic structure and behavior of the charge order. At low temperatures, they mediate competing lattice and electronic degrees of freedom accommodating local commensuration in a globally incommensurate structure. While most of the order remains statically locked in its initial configuration as the material is heated below the melting transition, the topological defects are remarkably mobile and reconfigure the surrounding order. Their motion suggests they are less strongly pinned to the lattice than the commensurate bulk. At high temperatures we find, in agreement with previous work, that the melting of the charge order is characterized by nucleation of additional discommensurations and dislocations, bringing about the concomitant shift in the ensemble wavevector and loss of long-range coherence observed through the transition. Intriguingly, as the order becomes heavily disrupted at high defect densities a growing proportion is also decoupled from the lattice periodicity becoming locally incommensurate.

These findings provide direct nanoscale insight into the evolution of incommensurability in charge ordered systems. They demonstrate the critical role of discommensurations in establishing incommensurate charge order [37], but also reveal the nanoscale coexistence of both locally commensurate and incommensurate periodicities as the order melts. Our results illustrate how nanoscale structural inhomogeneities cause charge order to unlock from the lattice periodicity, demonstrating that common physics established in the transition metal dichalcogenides, cuprates, and manganites governs incommensurability in both Fermi nesting driven charge density waves and the charge stripe ordering of doped holes strongly coupled to the atomic lattice [12,29,30,37,38,40,57].

This work demonstrates the promise of extending cryo-STEM nanocharacterization techniques to address further complexities within the phase diagrams of strongly correlated materials such as manganites. For instance, how does melting proceed in ½ hole-doped manganites, where charge ordered and ferromagnetic phases have been found to coexist and scattering measurements have shown that changes in wavevector generating incommensuration precede the loss of long-range order unlike the concomitant transitions seen in the 2/3 hole-doped system studied here [40,42,44,59]? Likewise, previous reports have shown scattering signatures of locally incommensurate charge order in the manganites characteristic of a uniform density wave – in these phases do topological defects still emerge and govern the evolution of the order, or is another mechanism at play [60]?

More broadly, our results illustrate the significance of nanoscale structural features in establishing the macroscopically observable properties of electronic order and governing its response to external stimuli. While in this study we considered a carefully selected model system, our observations of charge order dynamics bear strongly on the broad class of phenomena governed by competing electronic and lattice degrees of freedom, such as the complex pair density wave states identified in cuprates and exotic superconductors such as UTe$_2$, the confounding and elusive nature of charge order and a corresponding lattice response in emerging unconventional nickelate superconductors, and the plethora of charge density waves found in layered transition metal dichalcogenides [14,32,33,51,61–64]. Establishing the structural dynamics encountered at the boundaries of electronic phases may open new routes towards harnessing and applying them. Towards this goal, this work demonstrates a breakthrough in technique for characterizing the plethora of rich phases endemic to quantum materials at the nanoscale, in particular the capability of atomic resolution variable temperature cryo-STEM to measure dynamic behavior with sufficient spatial resolution and at the temperatures necessary to disentangle and stabilize phases of interest in these complex systems.

## Methods
### Sample preparation
Bi$_{1-x}$Sr$_{x-y}$Ca$_y$MnO$_3$ single crystals were grown with the flux method, using Bi$_2$O$_3$, CaCO$_3$, SrCO$_3$ and Mn$_2$O$_3$, and composition was determined to be approximately $x \approx 0.65$ and $y \approx 0.47$ with energy dispersive X-ray spectroscopy [43]. Thin cross-sectional lamellae in the (010) plane were lifted out with a Thermo Fisher Helios G4 UX focused ion beam (FIB) and attached to copper grids to minimize charging for thinning to electron transparency. Following this initial thinning, lamellae were cut off the copper grids and transferred to DENSSolutions heating/biasing Nano-Chips [65], where the lamellae were mounted with FIB-deposited platinum over holes cut in the silicon nitride membrane at the center of the Nano-Chip heating coil. After mounting to the Nano-Chips, the lamellae were re-thinned with the FIB to remove platinum and other material deposited over the region of interest during mounting and achieve a final thickness estimated to be 25-50 nm. Several lamellae lifted out from different regions of the single crystal were used for this and previous studies, differences in measured wavevectors indicate that stoichiometry variations on the order of 1% may be present, though all behavior presented here was observed in all lamellae. The data underlying Figures 1, 2, 3, 5 and 6 was collected from a single few-micron area of one lamella.

### Electron Microscopy
Selected area electron diffraction and atomic-resolution STEM imaging were performed on an aberration-corrected Thermo Fisher Titan Themis CryoS/TEM operated at 300 kV. High-angle annular dark-field imaging was performed with a 30 mrad convergence semi-angle, and 68 to 340 mrad collection angle. Sample temperature was controlled with a HennyZ FDCHB-6 continuously variable temperature double tilt liquid nitrogen sample holder [54] which supplied current to the Nano-Chips heating coil from an external power supply. Details on temperature control and calibration are found in the Supplementary Note 1 [55]. A constant holder rod heater temperature of 20 °C was used to minimize drift. Stacks of rapidly acquired frames (200-300 ns pixel dwell times) were acquired, rigidly registered [66], and averaged to reduce artifacts from drift and instability. Custom software and existing processing frameworks [43,56,67,68] were adapted to analyze the atomic resolution STEM data, as described in the Supplementary Notes 2 and 3 [55].

## Data Availability
The datasets generated and analyzed during the current study are available at the Platform for the Accelerated Realization, Analysis, and Discovery of Interface Materials (PARADIM) database, doi.org/XX.XXXXX.

## Acknowledgements

The authors thank David A. Muller, Suk Hyun Sung, Robert Hovden, and James L. Hart for helpful discussions and Elisabeth Bianco for experimental support. This work was primarily supported by the National Science Foundation (Platform for the Accelerated Realization, Analysis, and Discovery of Interface Materials (PARADIM)) under Cooperative Agreement No. DMR-2039380. The work at Rutgers University was supported by the DOE under Grant No. DOE: DE-FG02-07ER46382. N.S. acknowledges support from the NSF GRFP under award number DGE- 2139899. B.H.G acknowledges support from Schmidt Science Fellows in partnership with the Rhodes Trust and the Max Planck Society. The authors acknowledge the use of facilities and instrumentation supported by NSF through the Cornell University Materials Research Science and Engineering Center DMR-1719875, a Helios FIB supported by NSF (DMR-1539918), and FEI Titan Themis 300 acquired through NSF-MRI-1429155, with additional support from Cornell University, the Weill Institute and the Kavli Institute at Cornell.

## Author Contributions
I.E. and L.F.K. conceived the project. J.K. and S.-W.C. synthesized the samples. N.S. performed electron microscopy measurements. N.S., B.H.G., G.P., and I.E analyzed the electron microscopy data supervised by L.F.K. N.S., B.H.G., and I.E. wrote the manuscript with input from all authors.

## Competing Interests
The authors declare no financial or non-financial competing interests.